# Extreme Value Based Estimation of Critical Single Event Failure Probability

Gennady I. Zebrev, Artur M. Galimov, Rustem G. Useinov, Ivan A. Fateev

*Abstract* — A new survival probability function of ICs under space ion impact is proposed. Unlike the conventional approach, the function is based on the extreme value statistics which is relevant to the critical single event effects.

*Index Terms*— Single event effects, Hardness assurance, Cross section, Sensitive area, Critical failure.

## I. INTRODUCTION

Single event effects in microelectronics devices can be loosely divided into two groups: the critical events such as SEB SEGR, SEL etc., and the non-critical events as soft errors or SEU. The hardness assurance methodology for those two groups of effects is conceptually different. The relatively frequent soft errors are described in terms of the mean soft error rate (SER) on orbit, which is characterized by average SEU cross sections as functions of LET data $\sigma_0(\Lambda)$ [1]. Unlike the soft errors, where radiation hardness is defined by the average metrics, the rare by definition the critical effects are characterized by the survival probability function at a given fluence of incident particles [2]. Hardness assurance relies in this case on the criterion of a sufficient value of the survival probability on orbit which is determined by extreme events. This paper offers a method of survival probability estimation based on extreme statistics of the ion strikes in the sensitive area of an integrated circuit (IC).

## II. MATHEMATICAL BACKGROUND

An area of the IC potentially susceptible to Single Event Effects (SEE) is traditionally referred to as a sensitive area *A*. In practice, the value of *A* can be identified either by CAD simulations or by the laser screening testing [3]. The cross section value at high LETs (>60 MeV cm$^2$/mg) can also be used for approximate estimation of the sensitive area *A*.

A simple relation between a sensitive area *A* and the mean cross section of the critical event $\sigma_0(\Lambda)$ can be written as a function of LET

$$\sigma_0(\Lambda) = c(\Lambda) A \qquad (1)$$

where $c$ ($\leq 1$) is a dimensionless probability that an ion strike into the sensitive area causes a critical effect. Notice that $c$ and $A$ (as well as $\sigma_0$ and $A$) are the independent parameters. The former is an intensive technology-sensitive parameter (e.g., $c \cong 0$ for the Single Event Latchups (SEL) in the SOI ICs), while the latter is proportional to the geometrical sizes of the circuits or, of their vulnerable regions (e.g., CMOS memory).

Let us consider statistics of failures and ion strikes into the sensitive area *A*. A circuit survival probability after $k-1$ strikes is $(1-c)^{k-1}$, while the failure probability at the next *k-th* strike obeys so-called geometric distribution $p_k^{geom} = (1-c)^{k-1} c$ [4]. An expected value of the strike number to failure can be calculated as follows:

$$s = \sum_{k=1}^{\infty} k \, p_k^{geom} = \sum_{k=1}^{\infty} k (1-c)^{k-1} c = \frac{1}{c} = \frac{A}{\sigma_0} \geq 1. \qquad (2)$$

The variance of the strike number to the first failure is

$$\operatorname{var} s = \langle k^2 \rangle - s^2 = \frac{2-c}{c^2} - \frac{1}{c^2} = \frac{1-c}{c^2}. \qquad (3)$$

So the relative fluctuation of the strike number to failure becomes

$$\frac{\sqrt{\operatorname{var} s}}{s} = \sqrt{1-c}. \qquad (4)$$

This means that for actual values $c \ll 1$, the standard deviation of the strike number to failure is of order of its average value regardless of the strike statistics. This shows a fundamental property of the geometric distribution, that distinguishes it from, for example, the Poisson distribution, in which the relative role of fluctuations decreases with an increase in the statistics of events. In fact, this means that the notion of the cross section averaged even over a very extensive failure statistics is not a relevant for direct computation of survival probability due to a decisive role of the extreme events.

Actually, the experimental mean cross section is defined as an inverse of the mean fluence between failures $\langle \Phi \rangle$ (MTBF)

$$\sigma_0 = \frac{N}{\sum_{i=1}^{N} \Phi_i} = \frac{1}{\langle \Phi \rangle}, \qquad (5)$$

where $N$ is a failure statistics number. Meanwhile, information about the fluence-to-failure distribution is lost after such the averaging. Full information is contained in the cumulative failure-to-fluence distribution function $r(\Phi)$. This distribution function can be obtained experimentally and is often formally approximated, for example, by the Weibull function $r = \exp\left[-(\sigma_w \Phi)^\beta\right]$, where $\sigma_w$ и $\beta$ are the fitting parameters [5, 6]. The case $\beta = 1$ corresponds to the exponential distribution without memory for which $\sigma_0$ is defined as a unique physical cross section calculated as in (5)

$$r(\Phi) \cong e^{-\sigma_0 \Phi}. \qquad (6)$$

Notice that (6) is nothing but a simplest one-parametric interpolation of complex experimental data, which could be better approximated e.g., by formally more flexible two-parametric Weibull function (see Fig. 1).

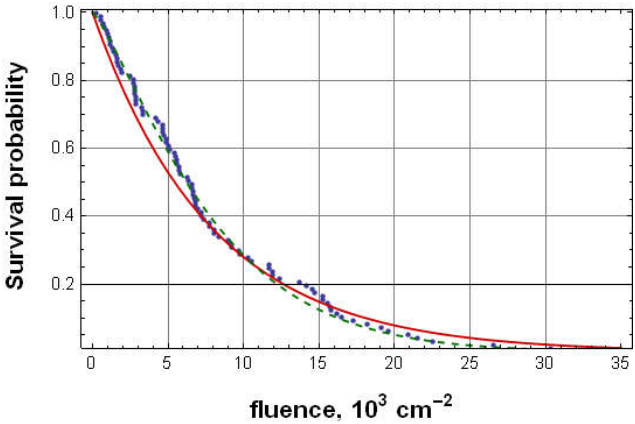

Fig. 1. A comparison of the experimental survival data (points) obtained from the controlled SEL statistics and the exponential distribution (solid line) with $\sigma_0 = 1.28 \times 10^{-4}$ cm$^2$ calculated via (5) and the Weibull distribution (dashed line) with fitted $\sigma_w = 1.2 \times 10^{-4}$ cm$^2$, $\beta = 1.25$.

The main issue is that both $\sigma_0$ and $r(\Phi)$ are not sufficient for the IC survival probability since the only events with the lowest fluence to failure are of concern [7]. In other words, unlike the SER prediction which is solely characterizing by mean value cross sections, the critical effect probability is defined by the extreme statistics.

The joint distribution that the minimum fluence-to-failure is greater than a given value $\Phi$ after $k$ strikes is $\left[r(\Phi)\right]^k$. Evidently, the actual strike number into the IC sensitive area in orbit can not be deterministically defined and only the expected value of this value is reasonable. Therefore, it should be averaged over the strike number statistics which is assumed to be the Poisson's distribution. As an absence of strikes ($k = 0$) certainly leads to zero effect, the averaging should be done over the zero-truncated Poisson distribution [8]:

$$\tilde{P}_k(z) = \frac{e^{-z}}{1-e^{-z}} \frac{z^k}{k!}, \quad \sum_{k=1}^{\infty} \tilde{P}_k(N) = 1, \qquad (7)$$

where $z = \Phi A$ is an expected value of the strike number. Note, unlike the integer $k \geq 1$, the expected value $z$ can take any real non-negative value $z \geq 0$. Then the new survival probability function has the following form:

$$R(\Phi) = \sum_{k=1}^{\infty} \left[r(\Phi)\right]^k \tilde{P}_k(\Phi A) = \frac{e^{\Phi A r(\Phi)} - 1}{e^{\Phi A} - 1} \qquad (8)$$

The same types of survival probability function are discussed in different contexts in [9, 10].

Fig. 2 shows a comparison of the extreme value based (8) and the mean value exponential survival functions calculated for the same cross sections. As can be seen from this figure, the extreme value approach gives generally a more conservative estimation of the survival probability.

Equation (8) has different asymptotic views in different ranges of $\Phi$:

$$R \cong \begin{cases} \exp(-\sigma_0 \Phi), & \Phi < A^{-1}, \quad (a) \\ \exp\left[-\Phi A\left(1 - e^{-\sigma_0 \Phi}\right)\right], & A^{-1} < \Phi < \sigma_0^{-1}, (b) \\ \exp(-A\Phi), & \sigma_0^{-1} < \Phi. \quad (c) \end{cases} \qquad (9)$$

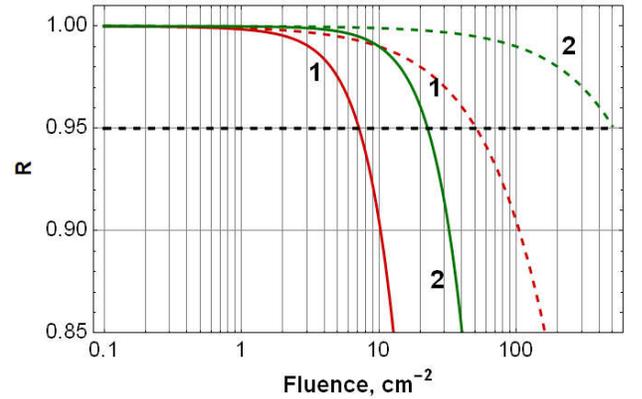

Fig. 2. The extreme value survival functions (solid lines), and the exponential survival functions (dashed lines) calculated as functions of fluence for $A = 1$ cm$^2$ and two cross sections (1) $\sigma_0 = 10^{-3}$ cm$^2$ and (2) $10^{-4}$ cm$^2$.

The first term (a) corresponds to extremely rare ion strikes into the IC sensitive area. The middle term (b) describes the case when the expected strike number greater than 1 but less than a mean strike number to failure $1 < k < s = 1/c$. Finally, the term (c) complies with the case of the abundant strikes, which provides a failure at any cross section value.

Fig. 3 shows the contour plot illustrating the extreme value function in the domain cross section–fluence. The curves separate the regions meeting the confidence level criterion of 0.95, 0.9 and 0.85.

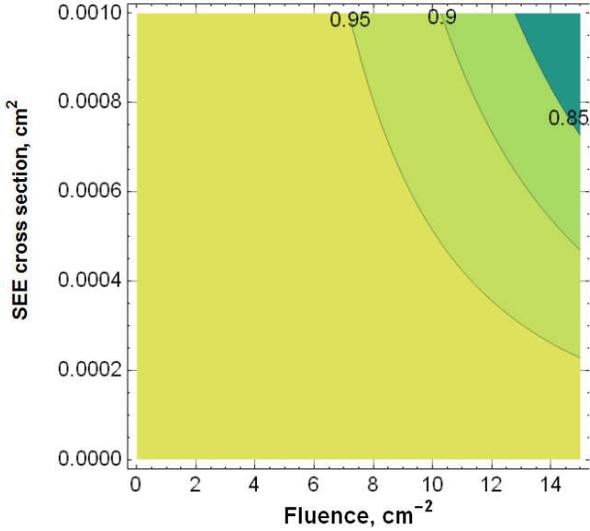

Fig. 3. The contour plot of the survival probability in the domain cross section – fluence, calculated for A = 1 cm². The curves define the regions with the confidence levels of 0.95, 0.9 and 0.85.

For example, the ten-year fluence in a geostationary orbit of heavy ions with LET in the range of 25 - 32 MeV-cm²/mg is of an order of 5 cm⁻². Then, as can be seen from Fig. 2, for the cross section $10^{-3}$ cm² such ions will not lead to failure for 10 years with the confidence level of 0.95.

### III. CASE STUDY AND APPLICATIONS

Equation (8) provides the survival probability at a given LET value. To estimate the reliability in space, the whole LET spectrum should be partitioned into $m$ bins in accordance with the available cross section data. Then, the generalized survival probability is the result of the multiplication of the partial survival probabilities:

$$R \cong \prod_{\lambda=1}^{m} R[\Phi_\lambda, \sigma_\lambda], \quad (10)$$

where index $\lambda$ distinguishes the LET intervals, $\sigma_\lambda$ and $\Phi_\lambda$ are the effective cross section and fluence in these intervals.

Equation (10) analysis allows determining the most "dangerous" LET ranges for a given IC in a given environment. Fig. 4 shows the cross section data for two hypothetical devices. The cross section vs LET dependencies can be conservatively approximated by the step functions.

The cross sections at low LETs (from 5 to 10 MeV-cm²/mg) were intentionally assumed the same, while the cross section for Dev.1 at 60 MeV-cm2/mg was chosen of an order of magnitude greater, than for Dev. 2. The sensitive area values are defined here as the cross sections at 60 MeV cm²/mg ($A_1 = 2\ 10^{-4}$ cm² and $A_2 = 3\ 10^{-5}$ cm²). Tables 1 and 2 show the contributions of different LET bins to the total probability calculated with different methods for the ten-year GEO spectrum for two devices. To a better comprehension, we illustrated the bin failure probabilities as 1 - $R_\lambda$ (see Fig. 5). For Dev. 1, the solid black line corresponds to the extreme value model; the dashed line corresponds to the classic model. Dev. 2 has the same legend style but the lines are red.

TABLE I SUMMARY OF THE CALCULATION RESULTS FOR DEV. 1

| LET range MeV-cm²/mg | $\Phi_\lambda$ cm⁻² | $A\Phi_\lambda$ | $\sigma_\lambda^{-1}$ cm⁻² | Classic $R_\lambda$ | Extreme $R_\lambda$ |
|---|---|---|---|---|---|
| 5-10 | 1.4 10⁴ | 2.8 | 2.5 10⁶ | 0.994 | 0.984 |
| 10-18 | 2.7 10³ | 0.5 | 2.0 10⁵ | 0.987 | 0.983 |
| 18-30 | 6.5 10² | 0.1 | 3.3 10⁴ | 0.980 | 0.979 |
| 30-45 | 2.8 | 5 10⁻⁴ | 1.0 10⁴ | 0.9997 | 0.9997 |
| 45-60 | 0.06 | 1 10⁻⁵ | 5 10³ = $A^{-1}$ | 0.99999 | 0.99999 |
| | | | | Total $R$ 0.959 | Total $R$ 0.927 |

TABLE II SUMMARY OF THE CALCULATION RESULTS FOR DEV. 2

| LET range MeV-cm²/mg | $\Phi_\lambda$ cm⁻² | $A\Phi_\lambda$ | $\sigma_\lambda^{-1}$ cm⁻² | Classic $R_\lambda$ | Extreme $R_\lambda$ |
|---|---|---|---|---|---|
| 5-10 | 1.4 10⁴ | 0.4 | 2.5 10⁶ | 0.994 | 0.993 |
| 10-18 | 2.7 10³ | 0.08 | 5.9 10⁵ | 0.995 | 0.994 |
| 18-30 | 6.5 10² | 0.02 | 1.4 10⁵ | 0.995 | 0.995 |
| 30-45 | 2.8 | 8 10⁻⁵ | 5.0 10⁴ | 0.9999 | 0.9999 |
| 45-60 | 0.06 | 2 10⁻⁶ | 3.3 10⁴=$A^{-1}$ | 0.99999 | 0.99999 |
| | | | | Total $R$ 0.982 | Total $R$ 0.979 |

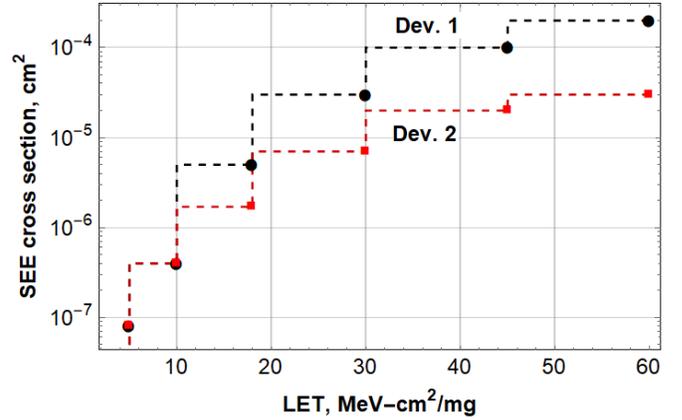

Fig. 4 The heavy ion cross section as function of LET for two hypothetical devices. The dashed lines show a conservative approximation used in calculations. The cross section for LET less than 5 MeV-cm²/mg is assumed to be zero.

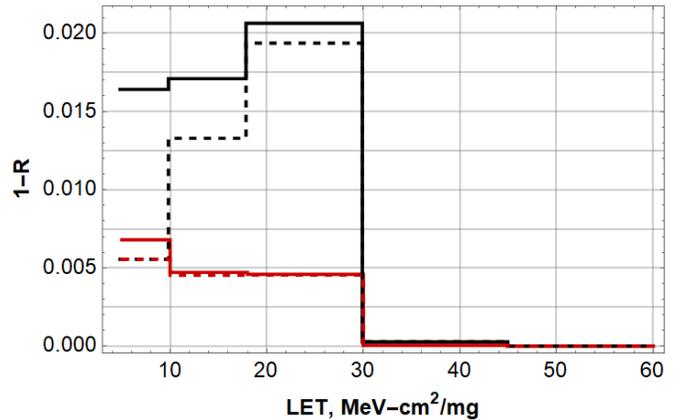

Fig. 5. Simulated failure probabilities in the LET ranges of cosmic rays spectrum on GEO for 10 years for cross section presented in Fig. 3. Black and red solid curves denote the results by the proposed model for Dev. 1 and 2 respectively. By analogy, the dashed curves denote the results by the classic model.

Fig. 5 and Tables 1 and 2 show an important role of the ion's strike statistics in the extreme failure probability calculations. In the LET range from 30 to 60 MeV-cm$^2$/mg, the expected strike numbers are very low ($A\Phi_\lambda$ <<1), and the failure probabilities are defined by only the cross sections as in the classic model (Eq. 9a). With increasing of fluence $\Phi_\lambda$, the value of $A$ becomes more important than the cross section. It can be seen from Fig. 5, the larger sensitive area of the device leads to the bigger difference between the extreme and classic failure probabilities. When the expected strike number becomes greater than 1, the failure probability is defined by the extreme asymptotic (Eq. 9b), which strongly depends on the $A$ value. The LET range from 5 to 10 represents this case for Dev. 1, where the failure probability is defined by the sensitive area value rather than cross section. Note, the classic model shows the same results for Dev. 1 and 2 in this range, that can be a potential source of the critical underestimation of the total failure probability.

If the laser screening showed larger sensitive areas for these devices, for example, $A_1 = 6 \times 10^{-4}$ cm$^2$ and $A_2 = 2 \times 10^{-4}$ cm$^2$, the total extreme $R$ would be 0.85 and 0.95 respectively that may be inappropriate for the system reliability. Meanwhile, the classic $R$ values remain the same as in Tables 1 and 2. This case highlights the importance of the additional testing of ICs on laser for the survival probability prediction in space.

## SUMMARY

To conclude, we highlight the next important points:
- The survival probability is determined not by average values, and by probability the extreme events with the lowest fluence-to-failure values.
- The extreme value model yields more conservative prediction.
- The IC sensitive area was found to be a significant independent parameter for prediction of failure probability which should be specified experimentally.
- The same cross section does not mean the same survival probability.
- The proposed model does not rely upon any particular physical mechanisms and can be applied to the critical soft errors as well.

## ACKNOWLEDGMENT

This work has been supported by FASIE under the "UMNIK" program contract #0038941 and the Competitiveness program of NRNU MEPHI.